\documentstyle[eqsecnum,aps]{revtex}
\begin{document}
\draft
\preprint{
ADP-00-29/T412\\
DTP-00/30}

\title{Does the weak coupling limit of the Burden-Tjiang
  deconstruction of the massless quenched QED3 vertex agree with perturbation 
theory?}

\author{A. Bashir\cite{byline}}

\address{Instituto de F{\'\i}sica y Matem\'aticas \\
Universidad Michoacana de San Nicol\'as de Hidalgo \\
Apdo. Postal 2-82 , Morelia, Michoac\'an, M\'exico}

\author{A. K{\i}z{\i}lers\"{u}\cite{byline}}
\address{Special Research Centre for the Subatomic Structure of 
Matter, \\
University of Adelaide, Adelaide, Australia 5005\\
and \\
Department of Physics and Mathematical Physics, \\
University of Adelaide, Adelaide, Australia 5005}

\author{M.R. Pennington\cite{byline}}
\address{Centre for Particle Theory,\\
University of Durham, Durham DH1 3LE, U.K.}
\date{\today}
\maketitle
\begin{abstract}
We derive constraints on the non-perturbative 3-point fermion-boson 
transverse vertex in  massless QED3 from its perturbative calculation to order
$\alpha$. We also check the transversality condition to
two loops and evaluate the fermion propagator to the same order.
We compare a conjecture of the non-perturbative vertex 
by Burden and Tjiang against our results and comment on its 
draw backs. Our calculation calls for the need to construct a 
non-perturbative form for the fermion-boson vertex which agrees with 
its perturbative limit to ${\cal O}(\alpha)$.
\end{abstract}

\section{INTRODUCTION}

QED in 3-dimensions (QED3) is a useful laboratory for studying the
strong coupling limit of a gauge theory. The lack of ultraviolet
divergences makes it easier to handle than its 4-dimensional
counterpart. Moreover, in the quenched approximation, it exhibits
confinement which makes it attractive for investigating strong
physics. The study of strong coupling gauge theories through the use
of Schwinger-Dyson equations requires knowledge of the
non-perturbative form of the fundamental fermion-boson interaction.
The most commonly used approximation is the bare vertex. However,
among other drawbacks, it fails to respect a key property of the
underlying field theory, namely the gauge invariance of physical
observables. An obvious reason is that the bare vertex fails to
respect the Ward-Green-Takahashi Identity (WGTI)~\cite{WGT}. Ball and
Chiu~\cite{BC} have proposed an {\em ansatz} for what is
conventionally called the longitudinal part of the vertex which alone
satisfies WGTI.  The rest of the vertex, the transverse part, remains
undetermined.  Dynamical fermion mass generation has been previously
studied in QED3, both quenched and unquenched, using the bare vertex,
as well as an {\em ansatz} based on a simple modification of the
Ball-Chiu vertex \cite{Roberts1, Walsh1}.  More recently, Burden and
Tjiang have constructed a different {\em ansatz} for the full vertex
to investigate fermion and photon propagators simultaneously
\cite{Burden1}, while including an explicit transverse piece. Burden
and Tjiang base their {\it deconstruction} on the assumption that a
certain \lq\lq transversality condition'' for the fermion propagator
holds non-perturbatively for some covariant gauge $\xi_0$.  The bare
fermion propagator is then a solution in that gauge. Accordingly, they
go on to propose a transverse vertex and use it to study the photon
propagator.

The only truncation of the complete set of Schwinger-Dyson equations
known so far that incorporates the key properties of a gauge theory at
each level of approximation is perturbation theory. Moreover, it is
natural to assume that physically meaningful solutions of the
Schwinger-Dyson equations must agree with perturbative results in the
weak coupling regime. While in QED4 this realization has been of
enormous help in constructing a physically acceptable form of the
vertex \cite{CP,KRP,BKP}, need exists to exploit perturbation theory
in exploring the non-perturbative form of the vertex in QED3.
Following \cite{KRP}, we evaluate the transverse part of the vertex to
${\cal O}(\alpha)$.  This result is then assumed to be the weak
coupling limit for the non-perturbative form of the transverse vertex.
We also check the Burden-Tjiang transversality condition to two loops
and find that to this order, it is not realized in perturbation
theory. We evaluate $F(p^2)$ to ${\cal O}(\alpha^2)$ analytically and
compare our findings with the conjecture of the vertex proposed by
Burden and Tjiang. 

\section{THE VERTEX}

The full vertex, Fig.~\ref{fig:fig1}, ${\Gamma^{\mu}(k,p)}$ can be expressed in
terms of 12 spin amplitudes formed from the vectors
${\gamma^{\mu},k^{\mu},p^{\mu}}$ and the scalars 1,${\not\!k,\not\!p}$
and ${\not\!k\not\!p}$.  It satisfies the Ward-Green-Takahashi
identity~\cite{WGT}
\begin{eqnarray}
q_{\mu}\Gamma^{\mu}(k,p)={\it S}^{-1}_{F}(k)-{\it S}^{-1}_{F}(p),
\end{eqnarray}
where ${q=k-p}$, and the Ward identity
\begin{eqnarray}
\Gamma^{\mu}(p,p)={\frac{\partial}{\partial p^{\mu}}}{\it S}^{-1}_{F}(p)
\end{eqnarray}
as the non-singular ${k \rightarrow p}$ limit of Eq.~(1). We follow Ball 
and Chiu and define the longitudinal component of the vertex in terms of
the fermion propagator as
\begin{eqnarray}
\Gamma^{\mu}_{L}(k,p)&=&\frac{\gamma^{\mu}}{2}
\left(\frac{1}{F(k^2)}+\frac{1}{F(p^2)}\right) \; + \; 
\frac{1}{2} \, \frac{({\not \! k}+{\not \! p})(k+p)^{\mu}}
{(k^2-p^2)}\left(\frac{1}{F(k^2)}-\frac{1}{F(p^2)}\right) \;.
\end{eqnarray}
This ${\Gamma^{\mu}_{L}}$ alone  satisfies 
the Ward-Green-Takahashi identity,
Eq.~(1), and being free of kinematic singularities the Ward identity,
Eq.~(2), too. The full vertex can then be written as
\begin{eqnarray}
\Gamma^{\mu}(k,p)=\Gamma^{\mu}_{L}(k,p)+\Gamma^{\mu}_{T}(k,p) \qquad,
\end{eqnarray}
where the transverse part satisfies
\begin{eqnarray}
q_{\mu}\Gamma^{\mu}_{T}(k,p)=0\;\;\;\;\;\mbox{and} \;\;\;\;
\Gamma^{\mu}_{T}(p,p)=0\qquad.
\end{eqnarray}
The Ward-Green-Takahashi identity fixes 4 coefficients of 
the 12 spin amplitudes
in terms of the fermion functions. The transverse component 
${\Gamma^{\mu}_{T}(k,p)}$
thus involves 8 vectors, of which the following 4 are sufficient to 
describe it in the chirally symmetric theory:
\begin{eqnarray}
\Gamma^{\mu}_{T}(k,p)=\sum_{i=2,3,6,8} \tau_{i}(k^2,p^2,q^2)T^{\mu}_{i}(k,p) 
\qquad,
\end{eqnarray}
where
\begin{eqnarray}
&T^{\mu}_{2}(k,p)&=\left[p^{\mu}(k\cdot q)-k^{\mu}(p\cdot q)\right]({\not\! k}
+{\not\! p})\nonumber\\
&T^{\mu}_{3}(k,p)&=q^2\gamma^{\mu}-q^{\mu}{\not \! q}\nonumber\\
&T^{\mu}_{6}(k,p)&=\gamma^{\mu}(p^2-k^2)+(p+k)^{\mu}{\not \! q}\nonumber\\
&T^{\mu}_{8}(k,p)&=-\gamma^{\mu}k^{\nu}p^{\lambda}{\sigma_{\nu\lambda}}
+k^{\mu}{\not \! p}-p^{\mu}{\not \! k}\nonumber\\
\mbox{with}\;\;\;\;\;\;\;\;\;
&\sigma_{\mu\nu}&=\frac{1}{2}[\gamma_{\mu},\gamma_{\nu}]\qquad.
\end{eqnarray}
The coefficients ${\tau_{\it i}}$ are Lorentz scalar functions of ${k}$ 
and ${p}$, i.e., functions of ${k^2,p^2,q^2}$. Solution of the 
Schwinger-Dyson equations for the fermion and photon propagators requires the
knowledge of ${\tau_{\it i}}$ in Eq.~(6). 

\subsection{Conjecture proposed by Burden and Tjiang}
Burden and Tjiang have recently proposed a non-perturbative
deconstruction of the  vertex \cite{Burden1} for massless QED3.
It involves certain assumptions about the fermion propagator and the
3-point fermion-boson vertex:

\subsubsection{The propagator and the transversality condition} 

In quenched QED, the SDE for the fermion propagator reads, Fig.~\ref{fig:fig2}:
\begin{eqnarray}
  i S_F^{-1}(p) &=& i {S_F^0}^{-1}(p) + e^2 \; \int \frac{d^3k}{(2 \pi)^3}
  \; \Gamma^{\mu}(k,p) \,  S_F(k) \, \gamma^{\nu} \, \Delta_{\mu \nu}^0(q)
  \qquad.
\end{eqnarray}
The photon propagator can be split into the transverse and the
longitudinal parts as:
\begin{eqnarray}
     \Delta_{\mu \nu}^0(q) &=&  {\Delta_{\mu \nu}^0}^T(q) - \xi \; 
     \frac{q_{\mu}q_{\nu}}{q^4} \qquad,
\end{eqnarray}
where
\begin{eqnarray}
      {\Delta_{\mu \nu}^0}^T(q) &=& -\frac{1}{q^2} \; \left[
 g_{\mu\nu}- q_{\mu}q_{\nu}/q^2\right] \qquad.
\end{eqnarray}
Burden and Roberts (see Eq.~(25) of \cite{trans}) have noted that the
solution of Eq.~(8)
is gauge covariant (in the sense of the Landau-Khalatnikov (LK) transformations
\cite{LK}) if the condition
\begin{eqnarray}
\int \frac{d^3k}{(2 \pi)^3}
  \; \Gamma^{\mu}(k,p) \,  S_F(k) \, \gamma^{\nu} \, 
  {\Delta_{\mu \nu}^0}^T(q)&=&0
\end{eqnarray}
is simply satisfied. This condition Burden and Tjiang \cite{Burden1}
have called the {\it transversality condition}. It is easy to check
that at one loop order this condition is indeed fulfilled and so we are left
with
\begin{eqnarray}
 i S_F^{-1}(p) &=& i {S_F^0}^{-1}(p) + e^2 \; \int \frac{d^3k}{(2 \pi)^3}
  \; \gamma^{\mu} \,  S_F^0(k) \, \gamma^{\nu} \, \left( - \xi \; 
     \frac{q_{\mu}q_{\nu}}{q^4} \right) \qquad.
\end{eqnarray}
Writing $S_F(p)={F(p^2)}/{\not \! p}$, in its most general form, the solution 
of the
above equation is:
\begin{eqnarray}
 F(p^2)&=&1-\frac{\alpha\xi}{4} \, \frac{\pi}{\sqrt{-p^2}} \; + \;
  {\cal O}({\alpha}^2) \quad. 
\end{eqnarray} 
At this point, it may be useful to compare this result with the implications
of the LK transformations and the expression proposed by Burden and Tjiang.
Assuming that $F(p^2)=1$ in the Landau gauge, LK transformations yield
the following expression for it in an arbitrary gauge: 
\begin{eqnarray}
F(p^2)&=&1-\frac{\alpha\xi}{2 \sqrt{-p^2}} \; 
{\rm tan}^{-1} \left[ \frac{2 \sqrt{-p^2}}{\alpha\xi} \right] \qquad.
\end{eqnarray}
Using the expansion ${\rm tan}^{-1}(1/x)= \pi/2 - x + 
x^3 /3 + \cdots$ for $ \mid x \mid << 1$, we  get
\begin{eqnarray}
   F(p^2)&=& 1 - \frac{\pi \, \alpha \xi}{4 \sqrt{-p^2}} 
- \frac{\alpha^2 \xi^2}{4p^2}
               + {\cal O}(\alpha^3) \quad,
\end{eqnarray}
which is in accordance with the perturbative result to ${\cal O}(\alpha)$.
Therefore, the LK transformations accompanied by the assumption that
$F(p^2)=1$ in the Landau gauge are in accordance with perturbation theory
at the one loop level. A similar comparison at the two loop level
is discussed in Sect. 4.

\noindent
Burden and Tjiang \cite{Burden1} propose the following
non-perturbative expression for $F(p^2)$:
\begin{eqnarray}
F(p^2)&=&1-\frac{\alpha(\xi - \xi_0)}{2 \sqrt{-p^2}} \; 
{\rm tan}^{-1} \left[ \frac{2 \sqrt{-p^2}}{\alpha(\xi-\xi_0)} \right] \qquad.
\end{eqnarray}
and they comment that \lq\lq Without knowing the transverse
contribution to $\overline{\Gamma}_{\mu}^{T}$ \footnote{$\gamma_{\mu}
  + \overline{\Gamma}_{\mu}^{T}$ is the solution of the SDE for the
  vertex when the fermion propagator is the bare propagator}, we are
unable to determine the constant $\xi_0$. The task of determining
$\overline{\Gamma}_{\mu}^{T}$ is a formidable task, and we have
nothing more to say about it in this paper."  However, it is trivial
to see, that as the weak coupling limit must agree with the
perturbative expansion, that $\xi_0=0$.

\subsubsection{The Burden-Tjiang vertex}

Burden {\it et al.} propose the following deconstruction of the vertex
in the Euclidean space:
\begin{eqnarray}
    \tau_i^{BT}(k^2,p^2)&=& \frac{1}{k^2-p^2} \; \left[ \frac{1}{F(k^2)} -
    \frac{1}{F(p^2)} \right] f_i(k^2,p^2) \hspace{30mm}   \\  \nonumber
{\rm where}  \hspace{28mm}   f_3(k^2,p^2)     
&=& \beta\; \frac{I(k,p)}{J(k,p)} \;, \\
    f_6(k^2,p^2)     &=& 0  \; , \\  
    {\overline{f}}(k^2,p^2) &=& -2 (1+ \beta)\, \frac{I(k,p)}{J(k,p)} \; ,
\end{eqnarray}
with
\begin{eqnarray*}
{\overline{\tau}}(k^2,p^2) &=&\tau_8(k^2,p^2)\,-\,(k^2+p^2)\, 
\tau_2(k^2,p^2) \hspace{65mm} \\
{\rm then} \hspace{32mm}   {\overline{f}}(k^2,p^2) &=& f_8(k^2,p^2) 
- (k^2 + p^2) f_2(k^2,p^2) \;, \\ \\
    I(k,p) &=& \frac{(k^2+p^2)^2}{8 k p} \, {\rm ln}\left| 
    \frac{k+p}{k-p} \right| - \frac{1}{4} \, (k^2+p^2) \;, \\  \\
     J(k,p) &=& \frac{(k^2-p^2)^2}{8 k p}\,  {\rm ln}\left| 
    \frac{k+p}{k-p} \right| - \frac{1}{4} \, (k^2+p^2)  \;.
\end{eqnarray*}
The superscript $BT$ in Eq.~(17) stands for Burden and Tjiang. Note a
few sign changes which had to be incorporated to rewrite their {\em
  ansatz} in terms of the basis $T^{\mu}_i$ that we have chosen in our
paper.  The form $(1/F(k^2) - 1/F(p^2))$ in Eq.~(17) has been chosen
to ensure that the transverse vertex vanishes in the gauge $\xi_0$
(note that we have shown that $\xi_0=0$).  Their vertex {\em ansatz}
is also based upon the assumption that the $\tau_i$ {\it have no $\xi$
  dependence other than a possible implicit dependence through F.}  In
order to see the validity of this {\em ansatz}, the following are some
of the important questions to be addressed:

\begin{itemize}

\item  Does the real transverse vertex vanish in the Landau gauge?

\item  Does perturbation theory allow us to take $\tau_6=0$, a coefficient
which plays a vital role in constructing the vertex in QED4?

\item  Does one loop perturbation theory agree with the non-perturbative
$\tau_i$ proposed by Burden {\it et al.}?

\item  For the selected basis $T_i$, do the corresponding coefficients
have kinematic singularities at the one loop level and beyond, as 
present in the {\em ansatz} of Burden {\it et al.} when $k \rightarrow p$?

\item Is $\beta$, which appears in the above {\em ansatz} independent of
the gauge parameter as claimed by Burden {\it et al.}?

\item  Does the {\it transversality condition}, Eq.~(11), hold true beyond 
the one loop order?

\end{itemize}
We carry out the one loop calculation of the vertex and the two loop 
calculation
of the fermion propagator to answer these questions.

\section{PERTURBATIVE CONSTRAINTS ON THE  VERTEX}

The vertex of Fig.~\ref{fig:fig1} can be expressed as
\begin{eqnarray}
\Gamma^{\mu}(k,p)=\,\gamma^{\mu}+\,\Lambda^{\mu}(k,p).
\end{eqnarray}

Using the Feynman rules, ${\Lambda^{\mu}}$ to
${O(\alpha)}$ is simply given by:
\begin{eqnarray}
-{\it i}e\Lambda^{\mu}(k,p)\,=\,\int_{M}\frac{d^3w}{(2\,\pi)^3}
(-{\it i}e\gamma^{\alpha}){\it i}{\it S}^{0}_{F}(p-w)(-{\it i}e\gamma^{\mu})
{\it i}{\it S}^{0}_{F}(k-w)(-{\it i}e\gamma^{\beta}){\it i}
\Delta^0_{\alpha\beta}(w) \;,
\end{eqnarray}
where $M$ denotes the loop integral is to be performed in Minkowski space.
The bare quantities are
\begin{eqnarray*}
  -{\it i}e\Gamma^0_{\mu}&=&-{\it i}e\gamma_{\mu}\\
  {\it i}{\it S}^0_{F}(p)&=&{\it i}
  {\not \! p}/{p^2}\\
  {\it i}{\it \Delta}^0_{\mu\nu}(p)&=&-{\it i}\left[p^2
 g_{\mu\nu}+(\xi-1)p_{\mu}p_{\nu}\right]/p^4  \quad,
\end{eqnarray*}
where $e$ is the usual QED coupling and the parameter ${\xi}$ specifies the
covariant gauge.
Following \cite{KRP}, ${\Lambda^{\mu}}$ can be re-expressed as:
\begin{eqnarray}
\Lambda^{\mu}(k,p)&=&-\frac{{\it i}\,{\alpha}}{2\,{\pi}^2}
\Bigg\{
\gamma^{\alpha}\,
{\not \! p}\,{\gamma^{\mu}}\,{\not \! k} \, \gamma_{\alpha}
{\bf {\it J}}^{(0)} 
-{\gamma^{\alpha}}\left(
{\not \! p}\,{\gamma^{\mu}}{\gamma^{\nu}}
+{\gamma^{\nu}}{\gamma^{\mu}}{\not \! k} \right)
\gamma_{\alpha}{\bf {{\it J}_{\nu}^{(1)}}}
+{\gamma^{\alpha}}{\gamma^{\nu}}{\gamma^{\mu}}{\gamma^{\lambda}}
\gamma_{\alpha}{\bf {{\it J}_{\nu\lambda}^{(2)}}}\nonumber\\
&& \hspace{8mm}
+(\xi-1)\Bigg[
\left(-{\gamma^{\nu}}{\not \! p}\,{\gamma^{\mu}}-\,
{\gamma^{\mu}}{\not \! k}\,{\gamma^{\nu}} \right)
{\bf {\it J}_{\nu}^{(1)}}
+{\gamma^{\mu}}{\bf \it K}^{(0)}
+{\gamma^{\nu}}{\not \! p}\,{\gamma^{\mu}}{\not \! k}\,{\gamma^{\lambda}}
 {\bf {{\it I}_{\nu\lambda}^{(2)}}}
\Bigg]
\Bigg\}, 
\end{eqnarray}
where ${\it J}^{(0)}$, ${\it J}^{(1)}_{\mu}$, ${\it J}^{(2)}_{\mu\nu}$,
${\it K}^{(0)}$ and ${\it I}^{(2)}_{\mu\nu}$ have been tabulated in the
appendix using the notation $k=\sqrt{-k^2}$, $p=\sqrt{-p^2}$,  
$q=\sqrt{-q^2}$. 
The only angular dependence
is displayed in $q= \sqrt{k^2 + p^2 - 2 kp \rm{cos \theta}}$.
The expression for the transverse vertex $\Gamma_T^{\mu}$ can be obtained
by subtracting from Eq.~(22), the contribution from the longitudinal part
$\Gamma_L^{\mu}$ at one loop. Eq.~(3) and Eq.~(13) allow us to write:
\begin{eqnarray}
    \Gamma^{\mu}_{L}(k,p) &=& \left[ 1 + \frac{\alpha \xi}{4} \, \eta_1 \right]
    \, \gamma^{\mu} \; + \;  \frac{\alpha \xi}{4} \, \eta_2 \, \left[ 
{k^{\mu}}{\not \! k} \, +  \, {p^{\mu}}{\not \! p} \, + \,
{k^{\mu}}{\not \! p} \, +  \, {p^{\mu}}{\not \! k}  \right] \qquad,
\end{eqnarray}
where
\begin{eqnarray}
   \eta_1 = \frac{\pi}{2} \, \left[ \frac{k+p}{kp}  \right] \qquad , \qquad
   \eta_2 = \frac{\pi}{2} \, \left[ \frac{1}{kp(k+p)}  \right] \quad ,
\end{eqnarray}

\begin{eqnarray}
  \tau_2(k^2,p^2) &=& \frac{\alpha \pi}{4} \; \frac{1}{kp(k+p)(k+p+q)^2} \;
 \left[  1 + (\xi-1) \, \frac{2k+2p+q}{q}  \right] \;, \\ \nonumber \\
   \tau_3(k^2,p^2) &=& \frac{\alpha \pi}{8} \; 
\frac{\left[ 4kp+3kq+3pq+2q^2 + (\xi-1) \, (2k^2+2p^2+kq+pq)\right] }
{kpq(k+p+q)^2} \;
  , \nonumber\\
   \\
 \tau_6(k^2,p^2) &=& \frac{\alpha \pi (2- \xi)}{8} \; \frac{k-p}{kp(k+p+q)^2} \;, \\ 
   \nonumber \\
 \tau_8(k^2,p^2) &=& \frac{\alpha \pi (2+ \xi)}{2} \; \frac{1}{kp(k+p+q)} \;.\\ \nonumber
\end{eqnarray}
  Any non-perturbative vertex {\em ansatz} should reproduce Eqs.~(25-28) 
in the 
weak coupling regime. Therefore, these equations should serve as 
a guide to constructing a non-perturbative vertex in QED3. Note that 
the ${\tau_i}$ have the required symmetry under the exchange of
vectors $k$ and $p$. ${\tau_2}$, ${\tau_3}$ and ${\tau_8}$ are symmetric,
whereas ${\tau_6}$ is antisymmetric.  All the $\tau_i$ only depend on 
elementary functions of $k$ and $p$. This
is unlike QED4, where the $\tau_i$ involve Spence functions.

Let us now try to answer some of the questions raised in the previous section:
\begin{itemize}
   
\item

   The transverse vertex does not vanish in the Landau gauge.

\item 

   The coefficient $\tau_6 \neq 0$. Moreover (as we shall see shortly), in  
the asymptotic limit  $k >> p$,  it contributes dominantly
to the transverse vertex along with $\tau_3$.

\item

   None of the $\tau_i$ agrees with the form proposed by Burden {\it et al.}
   The real $\tau_i$ are explicitly functions of $q^2$. However, we 
   shall later make a comparison with the proposed vertex in the key limit
for loop integrals
   when $k >> p$ where the real $\tau_i$ become independent of
   $q^2$, and a direct analogy with the proposed vertex is possible.

\item
   Very importantly, none of the ${\tau_i}$ has kinematic singularity when 
$k^2 \to p^2$.
One should note that {\it a priori} there was no guarantee that the set of 
basis
vectors $T_i$ which ensure their coefficients be independent of any 
kinematic
singularities in QED4 would achieve the same for QED3. However, we find
that these are indeed a correct choice for QED3 as well. As Burden {\it et al.}
realise the
logarithmic kinematical singularity in their vertex {\em ansatz} 
is, of course, ruled out by our perturbative calculation.

\end{itemize}
It is instructive to take the asymptotic limit $  k  >> 
 p  $ of the transverse 
vertex:
\begin{eqnarray}
    \tau_2(k^2,p^2)  &\stackrel{  k >>  p } {=}& - \frac{\alpha}{16k^4} \; 
           \frac{\pi}{p} \; (2-3 \xi) \; + \; {\cal O}(1/k^5)  \\
           \nonumber \\
    \tau_3(k^2,p^2)  &\stackrel{ k >>  p}{=}&  \frac{\alpha}{32k^2} \; 
           \frac{\pi}{p} \; (2+3 \xi) \; + \; {\cal O}(1/k^3)  \\
           \nonumber  \\
    \tau_6(k^2,p^2)  &\stackrel{ k >>  p}{=}&  \frac{\alpha}{32k^2} \; 
           \frac{\pi}{p} \; (2- \xi) \; + \; {\cal O}(1/k^3)  \\
           \nonumber \\ 
    \tau_8(k^2,p^2)  &\stackrel{ k >>  p  }{=}&  \frac{\alpha}{4k^2} \; 
           \frac{\pi}{p} \; (2+ \xi) \; + \; {\cal O}(1/k^3)  
           \;.
\end{eqnarray}
Note that taking into account the asymptotic limit  $ k  >>  p
 $ of the 
corresponding
basis vectors, one can easily see that $\tau_3$ and $\tau_6$
provide the dominant contribution to $\Gamma_T$ in this
limit just as in QED4. We are now in a position to compare
Eqs. (17-19) with Eqs. (29-32) in the limit $k >> p$ to try to extract
the value of $\beta$. Comparing $\tau_3$, we find
\begin{eqnarray}
   \beta= \frac{1}{2 \xi} \; + \; \frac{3}{4} \qquad ,    
\end{eqnarray}
which has an explicit dependence on $\xi$ contrary to the assumption of
Burden {\it et al.} Moreover, one could also extract the value of $\beta$
by comparing $\overline{\tau}$. Such an exercise leads us to
\begin{eqnarray}
   \beta= - \frac{5}{4} \; \left[ 1\;  + \; \frac{2}{\xi}  \right]   \qquad,
\end{eqnarray}
which is inconsistent with the value found earlier. Therefore, the
parametrization of the transverse vertex proposed by Burden and Tjiang
cannot be correct.

\section{$F(p^2)$ TO TWO LOOPS AND TRANSVERSALITY CONDITION}

\subsection{$F(p^2)$ to two loops}

We have seen that the transversality condition, i.e. Eq.~(11), holds
true to one loop level.  Burden {\it et al.} \cite{Burden1} have
proposed their vertex ansatz assuming this condition to be true
non-perturbatively for $\xi=\xi_0$, which we have shown must equal
zero.  Therefore, a crucial test of the validity of their vertex {\em
  ansatz} is checking the transversality condition to two loop order.
This is equivalent to calculating $F(p^2)$ to the same level. We carry
out this exercise in this section.

The equation for $F(p^2)$ can be extracted from Eq.~(8) by multiplying
the equation with ${\not \! p}$ and taking the trace. On Wick rotating
to the Euclidean space and simplifying, this equation can be written
as: 
\begin{eqnarray}
\nonumber
\frac{1}{F(p^2)}&=&1 \;-\; \frac{\alpha}{2 \pi^2 p^2} \; \int 
\frac{d^3k}{k^2} \frac{F(k^2)}{q^2}  \\ \nonumber
&&  \hspace{10mm} \Bigg[a(k^2,p^2) \, \frac{2}{q^2}  \, \left\{ ( k \cdot p )^2
 - (k^2+p^2) k \cdot p +k^2 p^2 \right\}    \\ \nonumber
&&    \hspace{13mm} + b(k^2,p^2)   \, \left\{ (k^2+p^2) k \cdot p +2 k^2 p^2
      - \frac{1}{q^2} (k^2-p^2)^2 k \cdot p  \right\} \\ \nonumber
&&  \hspace{13mm} - \frac{\xi}{F(p^2)} \frac{1}{q^2}   \, 
\left\{ p^2 (k^2 - k \cdot p) \right\} \\ \nonumber
&&  \hspace{13mm} + \tau_2(k,p) \; \;\left\{-(k^2+p^2) \Delta^2  \right\}
   \\ \nonumber
&&   \hspace{13mm} +  \tau_3(k,p)\, 2 \left\{- (k \cdot p)^2 
   + (k^2+p^2) k \cdot p - k^2 p^2  \right\}   \\ \nonumber
&&  \hspace{13mm} -  \tau_6(k,p) \, 2 \left\{ (k^2-p^2) k \cdot p  \right\} \\
&& \hspace{13mm} -  \tau_8(k,p) \; \; \,\left\{ \Delta^2 \right\}
\Bigg]   \qquad,
\end{eqnarray}
\baselineskip=6.8mm
\noindent where
\begin{eqnarray}
a(k^2,p^2)= \frac{1}{2} \, \left( \frac{1}{F(k^2)} + 
\frac{1}{F(p^2)} \right),
\hspace{7mm} 
b(k^2,p^2)= \frac{1}{2} \frac{1}{k^2-p^2} \, 
\left( \frac{1}{F(k^2)} - \frac{1}{F(p^2)} \right) \;.
\end{eqnarray}
Now using the expressions for
$\tau_i$, Eqs.~(25-28), we arrive at:
\begin{eqnarray}
\nonumber
\frac{1}{F(p^2)} &=& 1 + 
\frac{\pi \xi}{4 p} \, \alpha - \frac{\alpha^2}{4 p^2} 
    \int_0^{\infty} dk  \, \frac{1}{2kp(k+p)} \\ \nonumber
 && \hspace{-5mm}  \Bigg[ \frac{\xi}{2} \, (k^2-p^2) \, \left\{ 
   -(k^2-p^2)^2 I_4 + I_0 \right\}   \\ \nonumber
&& \hspace{-2mm} + \frac{\xi}{2} \, \left\{ (k^2-p^2)^2 (k^2+p^2) I_4 -
    2 (k^2+p^2)^2 I_2 + (k^2 + p^2) I_0 \right\}   \\ \nonumber
&& \hspace{-2mm} -\xi^2 p^2 (k^2-p^2)  \left\{ (k^2 - p^2) I_4 + I_2  \right\}
    \\ \nonumber
&& \hspace{-2mm} + \big\{ (k+p) \big(2kp(k-p)^2 I_3 - 3 (k-p)^2 (k+p) I_2
            +(3 (k-p)^2-2kp) I_1  \\ \nonumber
&& \hspace{-2mm} + 3(k+p) I_0 - 3 I_{-1} \big) \\ \nonumber
&& \hspace{1mm} +\xi \big( -kp(k-p)^2  I_2 + (k+p) (k^2+p^2) I_1
    +kp I_0 
  - (k+p) I_{-1} \big) \big\}    
      \Bigg]    \quad,
\end{eqnarray}
where:
\begin{itemize}

\item 
  
  The first curly-bracket expression arises from the $a$--term in
  Eq.~(35), the second one from the $b$--term, the third from the
  $\xi/F(p^2)$-term and the fourth from the transverse part of the
  vertex. On substituting $I_4$, $a$--term vanishes identically as it
  does at one loop level.  Note that all the $(k+p+q)$ factors in the
  $\tau_i$ neatly cancel out, leaving us with simpler integrals to be
  evaluated.

\item 

The $I_n$ are defined as
\begin{eqnarray*}
    I_n &=& \int_0^{\pi} \, d \theta \; \frac{{\rm sin} \theta}{q^n}
\end{eqnarray*}
with the evaluated expressions given in the appendix.
   
\end{itemize}
Keeping in mind the form of the integrals $I_n$, we divide the integration
region in two parts, $0 \rightarrow p$ and $p \rightarrow \infty$. For the
first region, we make the change of variables $k=px$ and for the
second region, $k=p/x$. On simplification, we arrive at
\begin{eqnarray}
\nonumber 
  \frac{1}{F(p^2)} &=& 1 + \frac{\pi \xi}{4 p} \alpha + 
  \frac{\alpha^2 \xi^2}{8 p^2} \int_0^1 \frac{dx}{x} \; \left[ 2 - (1-x)^2
   {\cal L} \right]  \\  \nonumber
 &&- \frac{\alpha^2}{8p^2} \int_0^1 \frac{dx}{x^2} (1-x) \left[
2(-2x^2+3x+3)-3(1-x)(1+x)^2  {\cal L}  \right] \\
 && - \frac{\alpha^2 \xi}{24 p^2} \int_0^1 \frac{dx}{x^2} \left[ 
   {2} (2x^3+5x^2+3x+3)-3 (x^2-x+1) (1+x)^2 {\cal L} \right] \,,
\end{eqnarray}

\noindent where
\begin{eqnarray}
       {\cal L} &=& \frac{1}{x} \; {\rm ln}\, \frac{1+x}{1-x} \quad.
\end{eqnarray}
The above integrals can be evaluated in a straightforward way.
In order to make a direct comparison with Eq.~(15), we
prefer to write the final expression in Minkowski space by
substituting  $p \rightarrow  \sqrt{-p^2}$  and $p^2 \rightarrow -p^2$:
\begin{eqnarray}
   F(p^2)&=& 1 - \frac{\pi \, \alpha \xi}{4 \sqrt{-p^2}}  - 
               \frac{\alpha^2 \xi^2}{4p^2}
              + \frac{3 \alpha^2}{4 p^2} \left(\frac{7}{3}-\frac{\pi^2}{4} 
              \right) +{\cal O}(\alpha^3) \,.
\end{eqnarray}
One can note various important features of this result:

\begin{itemize}

\item

    $F(p^2)  \neq 1$ in the Landau gauge. In fact, there is no value of the
covariant  gauge parameter $\xi$ for which $F(p^2)$ can be $1$.

\item 

 The existence of constant term at 
 ${\cal O}(\alpha^2)$ implies the violation
 of the transversality condition. We shall elaborate more on 
 this
  remark in Sect. 4.2.

\item 
  
  Eq.~(14) is derived from the LK transformations based upon the
  assumption that $F=1$ in the Landau gauge.  As we have seen, this
  assumption is not correct to ${\cal O}(\alpha^2)$, and therefore,
  Eq.~(14) is not expected to hold true in general, as is confirmed on
  comparing Eq.~(15) and Eq.~(39).  However, a comparison between the
  two results suggests that it contains the correct ${\cal O}(\xi^2)$
  term at the level ${\cal O}(\alpha^2)$, though it does not reproduce
  other term appearing in the exact perturbative calculation.

\end{itemize}

\subsection{Burden-Tjiang transversality condition}

The perturbative expression for $F(p^2)$ to the two loops
shows that the Burden-Tjiang transversality condition does not hold true
beyond one loop order. 
 Now we explicitly calculate the left hand side of Eq.~(11).
 In the most general form, it can be expanded as:
 \begin{eqnarray}
   i \,  \int \frac{d^3k}{(2 \pi)^3}
  \; \Gamma^{\mu}(k,p) \,  S_F(k) \, \gamma^{\nu} \, 
  {\Delta_{\mu \nu}^0}^T(q)&=& A(p^2) + B(p^2) \not \! p \quad,
\end{eqnarray}
where the multiplication with $i$ is only for mathematical convenience.
$A(p^2)$ and $B(p^2)$ can be extracted by taking the trace of the
above equation, having multiplied by $1$ and $\not \! p$ respectively. 
With the bare fermion being
massless, it is easy to see that on doing the trace algebra and
contracting the indices, $A(p^2)=0$. Our evaluation of $F(p^2)$ helps us
identify $B(p^2)$ from Eq.~(39) so that:
\begin{eqnarray}
\nonumber   && i \,  \int \frac{d^3k}{(2 \pi)^3}
  \; \gamma^{\mu} \,  S_F(k) \, \Gamma^{\nu}(k,p) \, 
  {\Delta_{\mu \nu}^0}^T(q)= \\
&& \hspace{20mm} \left[   
              - \frac{3 \alpha}{16 \pi p^2} \left(\frac{7}{3}
            - \frac{\pi^2}{4} \right)
             + {\cal O}(\alpha^2) \right] \; \not \! p  \;.
\end{eqnarray}
Obviously, for $\xi=0$,
\begin{eqnarray}
 i \,  \int \frac{d^3k}{(2 \pi)^3}
  \; \gamma^{\mu} \,  S_F(k) \, \Gamma^{\nu}(k,p) \, 
  {\Delta_{\mu \nu}^0}^T(q)  \mid_{\xi=0} &=&
            \left[   - \frac{3 \alpha}{16 \pi p^2} \left(\frac{7}{3} - 
              \frac{\pi^2}{4} 
                \right)+ {\cal O}(\alpha^2) \right]
               \; \not \! p  \;,
\end{eqnarray}
which is a violation of the transversality condition at the two
loop level.

\section{CONCLUSIONS}

In this paper, we have presented the one loop calculation of the
fermion-boson vertex in QED3 in an arbitrary covariant gauge for
massless fermions.  In the most general form, the vertex can be
written in terms of 12 independent Lorentz vectors.  Following the
procedure outlined by Ball and Chiu, 4 of the 12 vectors define the
longitudinal vertex.  It satisfies the Ward-Green-Takahashi identity
which relates it to the fermion propagator. The transverse vertex is
written in terms of the remaining 8 vectors. For massless fermions,
only 4 of these vectors contribute. Subtraction of the longitudinal
vertex from the full vertex yields the transverse vertex. We evaluate
the coefficients of the basis vectors for the transverse vectors to
${\cal O}(\alpha)$. Moreover, using this result, we calculate $F(p^2)$
analytically to ${\cal O}(\alpha^2)$ and find that the transversality
condition proposed by Burden and Tjiang does not hold true to this
order. Therefore, any non-perturbative construction of the transverse
vertex based upon this condition cannot be correct.

Knowing the vertex in any covariant gauge may give us an understanding
of how the essential gauge dependence of the vertex demanded by its
Landau-Khalatnikov transformation~\cite{trans,LK} is satisfied
non-perturbatively. Moreover, the perturbative knowledge of the
coefficients of the transverse vectors provides a reference for the
non-perturbative construction of the vertex as every {\em ansatz}
should reduce to this perturbative result in the weak coupling regime.
In comparison to the transverse vertex obtained by K{\i}z{\i}lers\"{u}
{\it et al.}  \cite{KRP} for QED4 (which contained Spence functions),
an important advantage of QED3 is that the corresponding results
contain only basic functions of momenta. This provides us with a
realistic possibility of searching for the non-perturbative form of
the transverse vertex.  The evaluation of $F(p^2)$ to ${\cal
  O}(\alpha^2)$ in an arbitrary covariant gauge should also serve as a
useful tool in the hunt for the non-perturbative vertex which is
connected to the former through Ward-Green-Takahashi identity and the
Schwinger-Dyson equations. Any vertex {\em ansatz} must reproduce
Eq.~(39) for $F(p^2)$ to ${\cal O}(\alpha^2)$ when the coupling is
weak, leading to a more reliable non-perturbative truncation of
Schwinger-Dyson equations: more reliable than the deconstruction of
Burden and Tjiang.

\acknowledgements
{}~~A.B. and A.K. are grateful for the hospitality offered to them 
by the Abdus Salam International Centre for Theoretical Physics (ICTP), 
Trieste, Italy for their stay there in the summer of 1998. M.R.P. and A.B. 
wish to thank the Special Centre for the Subatomic Structure of
Matter (CSSM), University of Adelaide, Adelaide, Australia, for the
hospitality extended by them during their respective visits of December 
1998--February 1999 and December 1999 to the Centre. We thank Andreas 
W. Schreiber for his useful comments on the draft version 
of the paper.

\section{APPENDIX}

Adopting the simplifying notation $k=\sqrt{-k^2}$, $p=\sqrt{-p^2}$ and
$q=\sqrt{-q^2}$, following are the some of the integrals used in the
calculation presented in the paper:
\begin{eqnarray}
 K^{(0)}&=&\int_{M}d^{3}w\, \frac{1}{(k-w)^2\,(p-w)^2} = 
\frac{i \pi^3}{q}  \\ \nonumber \\
 J^{(0)}&=& \int_{M}d^{3}w\,\frac{1}{ w^2 \, (p-w)^2\,
(k-w)^2} = \frac{-i \pi^3}{kpq}
   \\ \nonumber \\ 
{\it J}^{(1)}_{\mu}&=&\int_{M}\,d^3w\,\frac{w_{\mu}}
{w^2\,(p-w)^2\,(k-w)^2}  \nonumber \\
&=&  \frac{-i \pi^3}{kpq(k+p+q)} \;  \left[p k^{\mu} + k p^{\mu}  \right] \\
\nonumber \\
{\it J}^{(2)}_{\mu\nu}&=&\int_{M}\,d^3w\,\frac{w_{\mu}w_{\nu}}
{w^2\,(p-w)^2\,(k-w)^2} \nonumber \\
&=&  \frac{-i \pi^3}{2 kpq(k+p+q)^2} \;  \Big[-g^{\mu \nu} kpq(k+p+q) +
 k^{\mu} k^{\nu} p (k+2p+q)    \nonumber \\ 
 && \hspace{40mm} + p^{\mu} p^{\nu} k (2k+p+q) +
 ( k^{\mu} p^{\nu} +  p^{\mu} k^{\nu}) kp         \Big] \nonumber  \\ \\
{\it I}^{(2)}_{\mu\nu}&=&\int_{M}\,d^3w\,\frac{w_{\mu}w_{\nu}}
{w^4\,(p-w)^2\,(k-w)^2}  \nonumber \\
&=&  \frac{i \pi^3}{2 k^3 p^3 q(k+p+q)^2} \;  \Big[-g^{\mu \nu} k^2 p^2 
q(k+p+q) +
k^{\mu} k^{\nu} p^3 (2k+p+q) \nonumber \\ 
&& \hspace{40mm}  +  p^{\mu} p^{\nu} k^3 (k+2p+q) + 
 ( k^{\mu} p^{\nu} +  p^{\mu} k^{\nu}) k^2 p^2         \Big] \nonumber  \\ \\
I_{-1} &=& \frac{2}{3kp} \; \left[ p(3k^2 + p^2) \theta(k-p) + k
            (k^2 + 3 p^2)  \theta(p-k) \right]
            \\ \nonumber \\
I_0 &=& 2  \\ \nonumber \\
I_1 &=& \left[ \frac{2}{k} \theta(k-p) + \frac{2}{p} \theta(p-k) \right]
 \\ \nonumber \\
I_2 &=& \frac{1}{2kp} \; {\rm ln}\, \frac{(k+p)^2}{(k-p)^2}
\\ \nonumber \\
I_3 &=& \frac{2}{kp(k^2-p^2)} \left[ p \theta(k-p) - k \theta(p-k) \right]
\\ \nonumber \\
I_4 &=& \frac{2}{(k+p)^2 (k-p)^2}
\end{eqnarray}

\begin{figure}
  
  \caption{One loop correction to the vertex.}
  \label{fig:fig1}
\end{figure}

\begin{figure}
  
  \caption{Schwinger-Dyson equation for fermion propagator in quenched QED.}
  \label{fig:fig2}
\end{figure}

\begin{references}
\bibitem{WGT} J.C. Ward, Phys. Rev.  {\bf 78}, 182 (1950);\\
              H.S. Green, Proc. Phys. Soc. (London) {\bf A66}, 873 (1953);\\
              Y.Takahashi, Nuovo Cim.{\bf 6}, 371 (1957).
\bibitem{BC} J.S. Ball and T.-W. Chiu, Phys. Rev. {\bf D22}, 2542 (1980).
\bibitem{Roberts1} C.J. Burden and C.D. Roberts, Phys. Rev. {\bf D44}, 540
(1991).
\bibitem{Walsh1} D.C. Curtis, M.R. Pennington and D. Walsh, Phys. Lett.
{\bf B295}, 313 (1992).
\bibitem{Burden1} C.J. Burden and P.C. Tjiang, 
Phys Rev. {\bf D58,} 085019 (1998). 
\bibitem{CP} D.C. Curtis and M.R. Pennington, {\bf D42}, 4165 (1990).
\bibitem{KRP} A. K{\i}z{\i}lers\"u, M. Reenders and M.R. Pennington, Phys.
Rev. {\bf D52}, 1242 (1995).
\bibitem{BKP} A. Bashir, A. K{\i}z{\i}lers{\"u}
and M.R. Pennington, Phys. Rev. {\bf D57}, 1242 (1998).
\bibitem{trans} C.J. Burden and C.D. Roberts, Phys. Rev. {\bf D47}, 5581 
(1993).
\bibitem{LK} L.D. Landau and I.M. Khalatnikov, Zh. Eksp. Teor. Fiz. {\bf 29},
89 (1956)
[Sov. Phys. JETP, {\bf 2}, 69 (1956)]\\
B. Zumino, J. Math. Phys. {\bf 1}, 1 (1960);\\
\end{references}
\end{document}